# Allocentric Navigation is Computationally Universal

**Gualtiero Piccinini**
University of Missouri; Department of Philosophy; Columbia, MO; USA
piccininig@umsystem.edu
https://orcid.org/0000-0002-0825-9071

**Abstract.** I present three proofs that idealized neurocomputational architectures capable of navigation guided by allocentric maps with landmark structure can be computationally universal. Navigation may occur online, in the environment, or offline, in the animal's head. The first proof encodes the counters of a two-counter machine as positions of movable markers on orthogonal axes. The second directly simulates a one-tape Turing machine using a writable tape-path embedded in the map. The third replaces the globally designated path with a two-dimensional field of landmarks carrying only local predecessor/successor information. I establish universality in two senses: each class of navigation models is Turing-complete, and each contains a single machine that simulates every machine in its class. These constructions are mathematically close to classical graph-based models, including Kolmogorov–Uspensky machines, storage-modification machines, and graph Turing machines, so the bare computability results are unsurprising. My contribution is to reconstruct them explicitly within an allocentric cognitive-map architecture whose representational and computational primitives are drawn from empirical and theoretical work on spatial navigation and cognitive control. This shows that an allocentric navigation-based architecture can be computationally universal and opens the possibility of reconstructing aspects of biological cognition, including symbolic processing, in map-based terms.

**Acknowledgments.** This work was partially done on the traditional and ancestral lands of the Osage Nation in a region shaped by the histories of many other Indigenous peoples. I thank Carl Sachs for recently reminding me of Wells's work [1], which may have helped inspire this project. Martínez [2] may also have been a source of inspiration. Thanks to Jeff Dauer, Pawel Pachniewski, Waldemar Rohloff, Stephen Selesnick, and two anonymous referees for helpful feedback on the manuscript.

## 1. Introduction

I prove three sufficiency theorems showing that idealized navigation-based architectures with allocentric maps and landmark structure can be computationally universal. The underlying mathematical ideas are close to classical results from computability theory, especially work on counter machines, graph-based machines, and locally navigable memory structures. My contribution is to reconstruct those results explicitly within the framework of navigation in allocentric cognitive maps, a type of neurocomputational architecture independently motivated by empirical and theoretical neuroscience. I thereby show how established universal computational models can be reformulated within an allocentric navigation-based architecture.

I use “neurocomputational” to characterize the target architecture: its relevant components are neural and their operations are neural computations. I do not mean that the machine models below are artificial neural networks; neural computation need not be modeled using a neural-network formalism [3, 4]. Counter machines, Turing machines, and storage-modification machines are machine models in the same sense. Here I characterize the architecture only finely enough to establish its computational power, leaving its network organization and dynamics open (Section 7).

In contemporary neuroscience, allocentric cognitive maps and associated cognitive control are widely discussed representational and computational structures that organize knowledge relative to stable environmental features rather than momentary egocentric position [5-7]. Many animals use allocentric, landmark-anchored spatial representations in navigation [8]. In mammals, hippocampal activity can replay previously experienced trajectories and, in some cases, preplay trajectories later expressed during navigation [9, 10]. There is increasing evidence that hippocampal-entorhinal cognitive maps support functions beyond navigation, including memory, inference, and abstract relational knowledge, and that related map-like representations are present in other brain regions [11-14]. There is also evidence that hippocampal map-like activity is organized around action rather than wayfinding alone: in CA1, population activity was better predicted by the animal’s evolving action plan than by sensory or spatial variables themselves, suggesting that hippocampal activity can encode planned action sequences rather than location as such [15]. Prefrontal and orbitofrontal control systems also contribute to goal-directed navigation, route planning, detouring, and prospective or simulated navigation within cognitive maps [16-20]. This evidence raises the question of what representational and computational power allocentric navigation systems possess.

To answer this question, I formulate and prove a family of claims of the following form: an architecture that represents space in an allocentric coordinate system, represents landmarks at locations within that system, stores information at landmarks, can internally simulate navigation among those landmarks, and can internally simulate performing operations at those landmarks can compute any Turing-computable function under idealizations standard in computability theory.

While the primary target of this article is the representational and computational power of architectures capable of simulating navigation and action internally within cognitive maps, the navigation need not occur offline for the universality results to obtain. The sort of allocentric navigation investigated in this article may be entirely simulated in an animal’s head or it may occur partly or wholly within the animal’s environment. The extent to which navigation occurs online or offline makes a difference to how cognitively demanding the task is, because keeping track of all the needed information in one’s head requires more cognitive resources than storing some of that information in the environment. For the purposes of the universality results, however, that difference is immaterial so long as the same operations are available.

The benchmark is Turing universality in the standard sense established by Turing's analysis of effective calculability [21]. A convenient auxiliary benchmark is universality via counter or register machines, especially the simple two-counter models associated with Minsky and related register-machine formalisms [22-24]. The first proof simulates two-counter machines by using two movable markers on coordinate axes to encode unbounded counters. The second proof simulates standard one-tape Turing machines by embedding a writable tape-path in the allocentric map. The third proof is the most local: it still simulates ordinary Turing machines directly, but it dispenses with any globally indexed path available to the machine itself and uses only local directional information at neighboring landmarks.

Computing systems are standardly said to be universal in two senses — for instance, Minsky [23] employs both. In the first sense, a system or a class of systems is Turing-universal, or equivalently Turing-complete, when the systems in that class can compute every Turing-computable function. This is the sense in which the $\lambda$-calculus, tag systems, and elementary cellular automaton Rule 110 are said to be universal [25]. In the second sense, a single machine is universal for a class when, given an encoded description of any machine in the class together with an input, it computes what that machine computes. This is the sense in which universal Turing machines are universal [23, 26, 27]. Definitions 3.6 and 3.7 make both notions precise. Theorems 4.5, 5.4, and 6.5 establish the first for each of the three navigation models; Corollary 6.8 establishes the second.

The constructions are mathematically close to several classical graph-based machine models. Accordingly, the bare universality results are mathematically unsurprising. What is new here, as far as I know, is the explicit reconstruction of those results inside a navigation-based allocentric-map architecture of a kind that is independently motivated by neuroscientific evidence. The significance of this contribution is architectural and interdisciplinary: it shows that a neurocomputational architecture often associated primarily with spatial navigation can, under standard idealizations, realize universal computation.

This conclusion bears on a familiar line of criticism according to which nonclassical neurocomputational architectures cannot by themselves explain the productivity, systematicity, compositionality, or generality of cognition and therefore require supplementation by a classical symbolic architecture [e.g., 28, 29]. It also bears on contemporary debates over whether the available evidence warrants assigning representational functions to neural states [e.g., 30]. The present results show that navigation-based architectures have sufficient representational and computational power. This entails that such architectures can, in principle, realize arbitrarily complex computable transformations, including transformations exhibiting the kinds of productivity, systematicity, and compositionality often taken to require a classical symbolic architecture.

The same conclusion also bears on a second and opposite source of pressure. Radical enactivist, ecological, and dynamicist approaches emphasize embodied agent–environment coupling and have often been developed as alternatives to orthodox explanations in terms of internal computation and representation [31-38]. The present results do not vindicate a return to classical internal symbol manipulation. Allocentric navigation is itself embodied, action-oriented, dynamically controlled, and often environmentally embedded. What the results show, rather, is that even architectures situated in all these ways can naturally be analyzed in terms of representational and computational power [cf. 39]. Once an allocentric navigation system includes stable map-like states, landmark relations, stored item information, local transitions, and controllable movement or simulated movement through a map, its operations can, at the relevant level of analysis, be described as computations over representations and assessed for the class of transformations they can realize. Thus, the relevant contrast is not between classical symbolic computation and no computation or representation at all. A navigation-based architecture can be nonclassical, embodied, and dynamically controlled while still having the same computability-theoretic power as classical symbolic models.

Together, these points open the possibility of explaining higher cognition, including symbolic processing itself, in terms of cognitive map-based computation. Cognitive maps do not normally store and process symbols in any linguistic or language-like sense. What they store is information about the allocentric relations among locations, landmarks, and other items located within the map, together with operations that may be performed on such items. The present results show that those resources are sufficient for universal computation. Symbolic processing therefore need not be treated as architecturally basic but may instead be reconstrued in terms of map-like resources already known to exist in the brain.

At the same time, I do not argue that any biological system literally implements a Turing machine or any other classical computing device, whether universal or otherwise. Nor do I imply that implementing classical Turing-style computation within cognitive maps is required to explain cognition. Nor, finally, do I claim that a computational model by itself accounts for a cognitive process; it does not. The point is, rather, that a computational architecture often seen as capable at best of allegedly "low-level" cognitive tasks such as navigation, or perhaps too weak even for navigation without the support of allegedly more powerful symbolic architectures [e.g., 29] can, under standard idealizations, support universal computation. To that extent, cognitive map-based architectures are rich enough in principle to realize any computable cognitive function. Whether they in fact explain cognition, and if so whether they do it by classical or nonclassical means, is a separate empirical question.

The remainder of this article proceeds as follows: Section 2 outlines historical antecedents, Section 3 provides formal preliminaries and finitary conventions, Sections 4 through 6 present the three models and proofs of universality, Section 7 discusses the significance of the proofs, Section 8 discusses the proofs' scope, and Section 9 concludes.

## 2. Historical background

While the motivation for the present results stems from reflections on the relation between allocentric navigation and computational universality, the proofs have historical antecedents within computability theory. The most relevant predecessors are graph-based machines in which memory is not a linear tape but a mutable or navigable graph with a distinguished active location. The oldest example is Kolmogorov and Uspensky's graph-based conception of algorithmic computation. In this model, machines operate locally on graph-structured memory, and the resulting class of machines computes all partial recursive functions [40].

A second close predecessor is Schönhage's storage-modification machine. Here memory is represented by a mutable directed graph and computation proceeds by local pointer manipulations around an active node. This is mathematically very close to a navigation-based architecture in which an agent accesses and updates memory through local moves among landmarks [41].

A third close family is graph Turing machines. Ackerman and Freer formulate computation on vertex-labeled, edge-colored graphs and explicitly prove that ordinary Turing computation can be recovered inside that framework. Their model differs from the present one in being designed first and foremost as a graph-computation formalism rather than as a cognitive-map architecture, but it is relevant for any claim that graph-like spatial organization plus local rules can support universality [42].

There is also a neighboring literature on graph-walking automata and graph exploration. This work is relevant because it treats movement through a graph as a computational primitive and analyzes how much can be achieved when an automaton navigates an environment using only local information. This literature is also useful as a cautionary contrast: finite-state navigation by itself is generally not universal unless one adds unbounded memory resources [43, 44].

On the neurocognitive side, the allocentric-map framework used in this article is motivated by a large literature on hippocampal and entorhinal representations. The original cognitive-map program of O'Keefe and Nadel was linked to evidence of place cells, and subsequent work on place cells, grid cells, boundary cells, head-direction cells, object-vector cells, and related systems has made allocentric spatial coding a central part of contemporary theories of navigation and memory [5, 7, 45, 46].

As far as I have been able to determine, no prior source makes universality arguments explicit in the distinctive vocabulary of allocentric coordinates, landmarks, and navigation while simultaneously situating the architecture within the empirical cognitive-map literature. In this respect the results are mathematically unsurprising but conceptually novel and theoretically significant for the mind sciences.

The three models presented below may be summarized as follows. The counter-map model uses two movable marker positions on orthogonal axes to simulate a two-counter machine, thereby showing how spatial position alone can encode unbounded memory. The tape-path model uses writable labels on landmarks along a designated bi-infinite path to simulate an ordinary one-tape Turing machine, thereby identifying landmarks with memory cells. The local-layout model uses writable labels plus only local predecessor/successor cues in a two-dimensional field to simulate an ordinary one-tape Turing machine while staying closest to navigation among landmarks using local structure. The three models differ in which representational and operational resources the machine is granted: two axis-bound markers with zero-detection tests and no writable labels; writable labels together with a path whose adjacency relation is given as a primitive; and writable labels together with nothing but locally stored direction data, with no path object in the model at all.

## 3. Preliminaries and finitary conventions

Throughout, $\mathbb{N} = \{0,1,2,\dots\}$ and $\mathbb{Z}$ is the set of integers. For a function $f$ and arguments $u, v$ in its domain and codomain respectively, $f[u \mapsto v]$ denotes the function agreeing with $f$ everywhere except at $u$, where it takes the value $v$. Monus, or truncated subtraction, is written $\dot{-}$: $a \dot{-} b = a - b$ if $a \geq b$ and $a \dot{-} b = 0$ otherwise. Only the case $b = 1$ is used below.

**Definition 3.1 (Deterministic one-tape Turing machine).** A deterministic one-tape Turing machine is a tuple

$$M = (Q, \Gamma, \square, q_0, q_h, \delta),$$

where $Q$ is a finite set of states, $\Gamma$ is a finite tape alphabet, $\square \in \Gamma$ is the distinguished blank symbol, $q_0 \in Q$ is the start state, $q_h \in Q$ is the halting state, and

$$\delta: (Q \backslash \{q_h\}) \times \Gamma \to Q \times \Gamma \times \{L, R\}$$

is the transition function. Let $T_\Gamma = \{t: \mathbb{Z} \to \Gamma \mid t(i) = \square \text{ for all but finitely many } i \in \mathbb{Z}\}$ be the set of tape functions: the finite-support functions from cell indices to letters. A configuration of $M$ is a triple

$$(q, h, t) \in Q \times \mathbb{Z} \times T_\Gamma,$$

where $q$ is the current state, $h$ is the integer head position, and $t$ is the current tape function. Thus $h$ is not an element of $\Gamma$: $\Gamma$ is the tape alphabet, $h$ is an integer indexing a tape cell, and $t$ is one member of the set $T_\Gamma$, different configurations in general having different tape functions.

The one-step relation $\vdash_M$ on configurations is defined as follows. Let $a = t(h)$ and suppose $\delta(q, a) = (q', b, D)$. Then

$$(q, h, t) \vdash_M (q', h', t[h \mapsto b]),$$

where $h' = h - 1$ if $D = L$ and $h' = h + 1$ if $D = R$.

Note that $T_\Gamma$ is closed under single-cell updates, so $t[h \mapsto b] \in T_\Gamma$. Configurations with state $q_h$ have no $\vdash_M$-successor; every other configuration has exactly one. Hence $\vdash_M$ is a partial function and $M$ is deterministic. Write $\vdash_M^*$ for its reflexive-transitive closure.

**Definition 3.2 (CM2-programs and their semantics).** Let $Lab$ be a finite set of labels containing a distinguished start label $\ell_0$ and a distinguished halt label $\ell_h$. The set $Instr(Lab)$ of CM2-instructions over $Lab$ is generated by the grammar

$$Instr ::= inc(k, \ell) \mid jzdec(k, \ell, \ell'), \quad k \in \{1,2\}, \ell, \ell' \in Lab.$$

A CM2-program is a pair $P = (Lab, \iota)$, where $\iota: Lab \backslash \{\ell_h\} \to Instr(Lab)$ assigns exactly one instruction to each non-halting label. A CM2-configuration is a triple $(\ell, m, n) \in$ $Lab \times \mathbb{N} \times \mathbb{N}$, where $\ell$ is the current label and $m, n$ are the values of the two counters $c_1, c_2$. The one-step relation $\vdash_P$ is defined by cases on $\iota(\ell)$:

(i) $\iota(\ell) = inc(1, \ell')$: $(\ell, m, n) \vdash_P (\ell', m+1, n)$; (ii) $\iota(\ell) = inc(2, \ell')$: $(\ell, m, n) \vdash_P (\ell', m, n+1)$; (iii) $\iota(\ell) = jzdec(1, \ell', \ell'')$: $(\ell, 0, n) \vdash_P (\ell', 0, n)$ and $(\ell, m+1, n) \vdash_P (\ell'', m, n)$; (iv) $\iota(\ell) = jzdec(2, \ell', \ell'')$: $(\ell, m, 0) \vdash_P (\ell', m, 0)$ and $(\ell, m, n+1) \vdash_P (\ell'', m, n)$.

Configurations with label $\ell_h$ have no $\vdash_P$-successor; every other configuration has exactly one. Hence $\vdash_P$ is a partial function and every CM2-program is deterministic. The partial function $\phi_P: \mathbb{N} \rightharpoonup \mathbb{N}$ computed by $P$ is defined by: $\phi_P(a) = b$ if and only if $(\ell_0, a, 0) \vdash_P^* (\ell_h, b, n)$ for some $n \in \mathbb{N}$.

**Remark 3.3 (Universality of CM2 and input conventions).** Deterministic two-counter machines are a standard universal model of computation [22-24]. The input convention matters. Two counters suffice to simulate any Turing machine when inputs and outputs are exponentially coded, as $2^a$ and $2^{answer}$; with the identity convention they are strictly weaker, because a two-counter machine cannot compute $a \mapsto 2^a$ [47]. Nothing below depends on the choice of convention. Theorem 4.5 simulates CM2-programs instruction by instruction, so the counter-based navigation model inherits whatever functions CM2-programs compute under the chosen convention.

Two features of these standard results are used below. First, the simulations in question are step-for-step in the sense of Definition 3.5: each step of the simulated machine is mirrored by a finite block of two-counter instructions, and the simulated machine halts exactly when the simulating one does. Second, because the machines so simulated include the universal one-tape Turing machines, there is a CM2-program $P_U$ that, given a Gödel-coded description of an arbitrary one-tape Turing machine together with a coded input, computes what that machine computes on that input.

**Definition 3.4 (Navigation architecture and navigation model).** I use "navigation neurocomputational architecture" in the cognitive-neuroscientific sense: a system of computing components that represents locations, allocentric relations, landmarks, items stored at landmarks, movements or simulated movements among landmarks, and operations or simulated operations performed at landmarks.

A navigation model is the corresponding mathematical model: a finitely described machine whose configurations include a finite control state, a current map location or cursor position, and a finitely represented arrangement of markers or stored landmark labels. The three navigation models used below are defined explicitly in Definitions 4.1, 5.1, and 6.1, together with their instruction syntax and their one-step relations.

This terminology is meant to keep two issues apart. As navigation models, the machines below are ordinary effective machines. As neurocomputational architectures, they are redescriptions of computation in the vocabulary of allocentric maps and navigation.

**Definition 3.5 (Step-for-step simulation).** Let $S$ and $S'$ be deterministic machines with configuration sets $C$ and $C'$ and one-step relations $\vdash$ and $\Rightarrow$ respectively. An injection $E: C \rightarrow C'$ is a step-for-step simulation of $S$ in $S'$ if

(a) for every $c \in C$ that has a $\vdash$-successor $d$, there is $k \geq 1$ such that $E(c) \Rightarrow^k E(d)$ and no configuration reached from $E(c)$ in fewer than $k$ steps lies in the image of $E$; and
(b) $c$ is $\vdash$-terminal if and only if $E(c)$ is $\Rightarrow$-terminal.

The derivation in (a) is called a macro-step. When $k = 1$ for every $c$, the simulation is lock-step. Because both machines are deterministic, (a) and (b) jointly imply that for all $c, d \in C$: $c \vdash^* d$ and $d$ is terminal if and only if $E(c) \Rightarrow^* E(d)$ and $E(d)$ is terminal. Halting behavior, final control state, and final memory contents are therefore preserved in both directions.

**Definition 3.6 (Turing-completeness of a class).** A class $K$ of machines is Turing-complete, equivalently Turing-universal, if for every deterministic one-tape Turing machine $M$ there exist a machine $A \in K$ and a computable injection $E$ that is a step-for-step simulation of $M$ in $A$.

**Definition 3.7 (Universal machine for a class).** A machine $A^* \in K$ is universal for $K$ if there are a computable coding $\langle\cdot\rangle$ of the machines in $K$ and computable input and output codings such that, for every $A \in K$ and every input $w$ for $A$, $A^*$ on input $\langle A\rangle$ followed by $w$ halts if and only if $A$ halts on $w$, and then yields the encoded output of $A$ on $w$.

Definitions 3.6 and 3.7 are logically independent in general, and both are called universality in the literature. Both are established below: Theorems 4.5, 5.4, and 6.5 give Turing-completeness for the three classes; Corollary 6.8 gives a universal machine within each class.

Note that when $K$ is Turing-complete, a machine universal for $K$ is thereby able to compute every Turing-computable function; the two notions are independent in general but coincide in that respect here.

**Convention 3.8 (Finite descriptions and finite support).** All machines considered in this article are finitely described. Their control sets and transition tables are finite. Their maps are unbounded only in the ordinary potential-infinity sense in which a Turing tape is unbounded. At any moment, only finitely many markers or nonblank landmark labels need to be represented explicitly. Sets such as $T_\Gamma$, and the set $\Lambda_\Gamma$ of landmark labelings introduced in Definition 5.1, are sets of finite-support functions; their members are metatheoretic descriptions of finite memory states, not completed infinite objects made available to the machine as oracles. This is the standard finitary interpretation of effective computation [21, 48].

With this convention in place, the theorems below establish ordinary Turing universality. The reverse direction, that each of the navigation models is itself simulable by a Turing machine, is verified in Corollary 6.7.

## 4. Universality by navigation on two coordinate axes

The first navigation model uses two one-dimensional axes embedded in a two-dimensional allocentric frame. One movable marker $X$ is constrained to the horizontal axis and one movable marker $Y$ is constrained to the vertical axis. Their distances from the origin encode the values of two unbounded counters. The internal control state of the machine plays the role of the finite control of a counter machine. Thus, the model does not require a rich two-dimensional field of memory locations; it uses two spatial registers distinguished by their allocentric axes.

**Definition 4.1 (Counter-based allocentric navigation machine).** Let $D = \{E, W, N, S\}$ be the set of directions and $B = \{atO, atX, atY\}$ the set of tests. For a finite set $Q$, the set of navigation instructions over $Q$ is generated by the grammar

$$NInstr(Q) ::= move(d, q) \mid shift(d, q) \mid test(\beta, q, q'), \quad d \in D, \beta \in B, q, q' \in Q.$$

A counter-based allocentric navigation machine is a tuple

$$A = (\mathbb{N}^2, O, Q, q_0, q_h, \Pi),$$

where $\mathbb{N}^2$ is the map, $O = (0,0)$ is the origin, $Q$ is a finite set of control states, $q_0 \in Q$ is the start state, $q_h \in Q$ is the halt state, and

$$\Pi : Q \backslash \{q_h\} \to NInstr(Q)$$

is the navigation program, which assigns exactly one instruction to each non-halting control state. Control states thus play the role of instruction labels, and a program is a finite labeled instruction list.

Two things should be kept apart here. The navigation program $\Pi$ is an aspect of the machine, in the same way that the transition table $\delta$ of a Turing machine is an aspect of that machine: it is fixed structure, not an object stored at locations in the map and read off during computation. Calling $\Pi$ a program therefore imports no stored-program or von Neumann organization — an ordinary Turing machine is not stored-program either. A fortiori, nothing in the models below requires that a navigation architecture store, address, or interpret encoded instructions. Corollary 6.8 does exhibit a navigation machine that reads an encoded description of another machine from the map itself; but that is a result about one thing that such architectures *can* do, not a claim that biological navigation is generally organized that way.

A configuration of $A$ is a quadruple

$$\gamma = (q, p, x, y) \in Q \times \mathbb{N}^2 \times \mathbb{N} \times \mathbb{N},$$

where $q$ is the control state, $p$ is the position of the cursor, and $x$ and $y$ are the positions of the two markers: marker $X$ occupies $(x, 0)$ and marker $Y$ occupies $(0, y)$. The cursor and the two markers are three distinct components of a configuration. The cursor is where the machine currently is, or simulates being; the markers are items located in the map. Marker positions are not projections of the cursor position, and neither determines the other.

**Definition 4.2 (Primitive operations and one-step relation).** The four move functions on $\mathbb{N}^2$ are total:

$$E(a, b) = (a + 1, b), \quad W(a, b) = (a \dot{-} 1, b), \quad N(a, b) = (a, b + 1), \quad S(a, b) = (a, b \dot{-} 1).$$

Using monus makes $W$ and $S$ total: at the corresponding axis boundary they leave the cursor in place. The three tests are predicates on configurations:

$$atO(\gamma) \Leftrightarrow p = O; \quad atX(\gamma) \Leftrightarrow p = (x, 0); \quad atY(\gamma) \Leftrightarrow p = (0, y).$$

The one-step relation $\Rightarrow_A$ is defined by cases on $\Pi(q)$:

(i) $\Pi(q) = move(d, q')$: $(q, p, x, y) \Rightarrow_A (q', d(p), x, y)$. (ii) $\Pi(q) = test(\beta, q', q'')$: $(q, p, x, y) \Rightarrow_A (q', p, x, y)$ if $\beta(q, p, x, y)$ holds, and $(q, p, x, y) \Rightarrow_A (q'', p, x, y)$ otherwise. Every conditional instruction thus has both a true branch and an explicit false branch. (iii) $\Pi(q) = shift(d, q')$. The direction of a shift determines which marker it affects: $E$ and $W$ shift $X$, while $N$ and $S$ shift $Y$. A shift takes effect only if the cursor coincides with the marker in question. For each direction $d$, exactly one of the following five sub-cases applies:

– $d = E$: $(q, (x, 0), x, y) \Rightarrow_A (q', (x, 0), x + 1, y)$;

– $d = W$: $(q, (x, 0), x, y) \Rightarrow_A (q', (x, 0), x \dot{-} 1, y)$;

– $d = N$: $(q, (0, y), x, y) \Rightarrow_A (q', (0, y), x, y + 1)$;

– $d = S$: $(q, (0, y), x, y) \Rightarrow_A (q', (0, y), x, y \dot{-} 1)$;

– in every other case, $(q, p, x, y) \Rightarrow_A (q', p, x, y)$.

Configurations with control state $q_h$ have no $\Rightarrow_A$-successor; every other configuration has exactly one, so $\Rightarrow_A$ is a partial function and $A$ is deterministic.

Clause (iii) fires only when the cursor already occupies the marker's location, and that requirement is the model's locality constraint: markers are items at locations, and the machine can act on an item only where the item is. Dropping the constraint would turn the model into an ordinary register machine with random access to its registers and would remove precisely what makes it a navigation model. Because of the locality constraint, a CM2-instruction is not translated by a single primitive operation but by a macro that travels to the relevant marker, acts on it, and returns to the origin.

**Definition 4.3 (Translation of CM2-programs into navigation programs).** Let $P = (Lab, \iota)$ be a CM2-program. Define the counter-based allocentric navigation machine

$$A_P = (\mathbb{N}^2, O, Q_P, q_{\ell_0}, q_{\ell_h}, \Pi_P),$$

where $Q_P$ consists of one control state $q_\ell$ for each $\ell \in Lab$ together with at most five auxiliary states $\ell^1, \dots, \ell^5$ for each $\ell \in Lab \backslash \{\ell_h\}$, all distinct. The translation $\Pi_P$ is defined by cases on $\iota(\ell)$. The four instruction forms are handled as follows:

(i) $\iota(\ell) = inc(1, \ell')$:

$$\Pi_P(q_\ell) = test(atX, \ell^1, \ell^2) \quad \Pi_P(\ell^2) = move(E, q_\ell)$$

$$\Pi_P(\ell^1) = shift(E, \ell^3) \quad \Pi_P(\ell^3) = test(atO, q_{\ell'}, \ell^4) \quad \Pi_P(\ell^4) = move(W, \ell^3)$$

(ii) $\iota(\ell) = inc(2, \ell')$: as in (i), with $atY$ for $atX$, $N$ for $E$, and $S$ for $W$.

(iii) $\iota(\ell) = jzdec(1, \ell', \ell'')$:

$$\Pi_P(q_\ell) = test(atX, q_{\ell'}, \ell^1) \quad \Pi_P(\ell^1) = move(E, \ell^2) \quad \Pi_P(\ell^2) = test(atX, \ell^3, \ell^1)$$

$$\Pi_P(\ell^3) = shift(W, \ell^4) \quad \Pi_P(\ell^4) = test(atO, q_{\ell''}, \ell^5) \quad \Pi_P(\ell^5) = move(W, \ell^4)$$

(iv) $\iota(\ell) = jzdec(2, \ell', \ell'')$: as in (iii), with $atY$ for $atX$, $N$ for $E$, and $S$ for $W$.

The map $P \mapsto A_P$ is computable, indeed given by a finite table. Its intended reading is navigational. In case (i) the machine walks east from the origin until it reaches $X$, shifts $X$ one step further east, walks back west to the origin, and enters the state corresponding to the next CM2-label. In case (iii) it first checks, without moving, whether $X$ is at the origin, which is the zero test for $c_1$; if it is not, the machine walks east to $X$, shifts $X$ one step west, and walks back to the origin. The zero test is a test on the simulating machine's configuration, not an operation of the simulated CM2-program, whose instruction jzdec has no origin to check.

Each of the four cases assigns a finite set of labeled instructions to the control state $q_\ell$ and its auxiliary states; that set is the *macro* for the instruction $\iota(\ell)$, and the program $\Pi_P$ is the union of the macros over all non-halting labels.

**Lemma 4.4 (Traversal).** In any counter-based allocentric navigation machine, move and test instructions leave the marker components $x$ and $y$ unchanged; only shift instructions change them. Consequently, for any control states involved:
(a) from a configuration with cursor at $(k, 0)$, where $k \leq x$, iterated eastward moves place the cursor at $(x, 0)$ after exactly $x - k$ moves; $atX$ is false at every earlier point of the walk and true then;
(b) from a configuration with cursor at $(k, 0)$, iterated westward moves place the cursor at $O$ after exactly $k$ moves; $atO$ is false at every earlier point and true then;
(c) the analogues of (a) and (b) hold on the vertical axis: from a configuration with cursor at $(0, k)$, where $k \leq y$, iterated northward moves place the cursor at $(0, y)$ after exactly $y - k$ moves, with $atY$ false at every earlier point of the walk and true then; and from a configuration with cursor at $(0, k)$, iterated southward moves place the cursor at $O$ after exactly $k$ moves, with $atO$ false at every earlier point and true then.

*Proof.* Immediate from Definition 4.2. Clauses (i) and (ii) of the one-step relation leave $x$ and $y$ untouched. Since $E(a, b) = (a + 1, b)$, after $j$ eastward moves from $(k, 0)$ the cursor is at $(k + j, 0)$, and $atX$ holds exactly when $k + j = x$. Since $W(a, b) = (a \dot{-} 1, b)$, after $j \leq k$ westward moves from $(k, 0)$ the cursor is at $(k - j, 0)$, and $atO$ holds exactly when $j = k$. ■

**Theorem 4.5.** For every CM2-program $P$, the machine $A_P$ of Definition 4.3 simulates $P$ step for step under the encoding

$$E_P(\ell, m, n) = (q_\ell, O, m, n).$$

Consequently, the class of counter-based allocentric navigation machines is Turing-complete.

*Proof.* $E_P$ is injective, since distinct CM2-configurations differ in a component that $E_P$ preserves. Condition (b) of Definition 3.5 holds because $\Pi_P$ is undefined exactly at $q_{\ell_h}$ and $\vdash_P$ is undefined exactly at configurations labeled $\ell_h$. For condition (a), fix $\ell \neq \ell_h$ and argue by cases on $\iota(\ell)$, starting from $E_P(\ell, m, n) = (q_\ell, O, m, n)$.

Case $\iota(\ell) = inc(1, \ell')$. The instruction at $q_\ell$ tests $atX$ with the cursor at $O$, so $atX$ holds if and only if $m = 0$. If $m = 0$, the machine enters $\ell^1$, shifts $X$ east (permitted, since the cursor coincides with $X$ at $O$), reaching $(\ell^3, O, 1, n)$; the test at $\ell^3$ succeeds and the machine enters $q_{\ell'}$, giving $(q_{\ell'}, O, 1, n) = E_P(\ell', 1, n)$. If $m > 0$, then by Lemma 4.4(a) the alternation of tests at $q_\ell$ and moves at $\ell^2$ brings the cursor to $(m, 0)$ after exactly $m$ moves, with $x$ and $y$ unchanged; $atX$ then holds, the machine enters $\ell^1$ and shifts $X$ east, reaching $(\ell^3, (m, 0), m + 1, n)$; by Lemma 4.4(b) the alternation of tests at $\ell^3$ and moves at $\ell^4$ returns the cursor to $O$ after exactly $m$ moves; the test at $\ell^3$ then succeeds and the machine enters $q_{\ell'}$, giving $E_P(\ell', m + 1, n)$. Either way the resulting configuration is the encoding of the $\vdash_P$-successor, and the total number of primitive steps is $4m + 3$. No intermediate configuration lies in the image of $E_P$. During the outward loop, a configuration with a nonauxiliary control state has the cursor away from the origin; every other intermediate configuration has an auxiliary control state. During the return loop, all intermediate configurations have auxiliary states.

Case $\iota(\ell) = jzdec(1, \ell', \ell'')$. The test at $q_\ell$ with the cursor at $O$ succeeds if and only if $m = 0$. If $m = 0$, the machine enters $q_{\ell'}$ without moving anything, giving $(q_{\ell'}, O, 0, n) = E_P(\ell', 0, n)$, which is the $\vdash_P$-successor by Definition 3.2(iii). If $m > 0$, the machine enters $\ell^1$ and, by Lemma 4.4(a) applied to the loop $\ell^1$–$\ell^2$, reaches $(\ell^3, (m, 0), m, n)$ after exactly $m$ moves and $m$ tests; the shift at $\ell^3$ moves $X$ one step west, giving $(\ell^4, (m, 0), m - 1, n)$; by Lemma 4.4(b) the loop $\ell^4$–$\ell^5$ returns the cursor to $O$, and the machine enters $q_{\ell''}$, giving $E_P(\ell'', m - 1, n)$, which is again the $\vdash_P$-successor. The same reasoning excludes intermediate encoded configurations.

Cases (ii) and (iv) are the images of (i) and (iii) under the exchange of axes, and the verification is identical with $atY, N$, and $S$ in place of $atX, E$, and $W$.

Hence $E_P$ is a step-for-step simulation. By Remark 3.3, for every one-tape Turing machine $M$ there is a CM2-program $P_M$ that simulates $M$ step for step in the sense of Definition 3.5; composing the two simulations, and noting that the composition of computable injections satisfying Definition 3.5 again satisfies it, yields a counter-based allocentric navigation machine simulating $M$. So the class is Turing-complete in the sense of Definition 3.6. ■

**Example 4.6 (How the model computes).** The following exhibits the primitive operations of Definition 4.2 in action.

Let $P_{succ} = (\{\ell_0, \ell_h\}, \iota)$ with $\iota(\ell_0) = inc(1, \ell_h)$, so that $\phi_{P_{succ}}(a) = a + 1$. Definition 4.3(i) yields $A_{P_{succ}}$ with control states $q_{\ell_0}, q_{\ell_h}, \ell_0^1, \ell_0^2, \ell_0^3, \ell_0^4$ and program

$$\Pi(q_{\ell_0}) = test(atX, \ell_0^1, \ell_0^2), \quad \Pi(\ell_0^2) = move(E, q_{\ell_0}), \quad \Pi(\ell_0^1) = shift(E, \ell_0^3), \quad \Pi(\ell_0^3) = test(atO, q_{\ell_h}, \ell_0^4), \quad \Pi(\ell_0^4) = move(W, \ell_0^3).$$

The computation on input $a = 2$ begins at $E_P(\ell_0, 2,0) = \left(q_{\ell_0}, (0,0), 2,0\right)$ and proceeds as shown in Table 1:

*Table 1 Primitive navigation steps in the successor example on input 2*

| Step | Instruction executed | Resulting configuration |
|---|---|---|
| 1 | $test(atX, \ell_0^1, \ell_0^2)$, $atX$ false | $(\ell_0^2, (0,0), 2,0)$ |
| 2 | $move(E, q_{\ell_0})$ | $(q_{\ell_0}, (1,0), 2,0)$ |
| 3 | $test(atX, \ldots)$, $atX$ false | $(\ell_0^2, (1,0), 2,0)$ |
| 4 | $move(E, q_{\ell_0})$ | $(q_{\ell_0}, (2,0), 2,0)$ |
| 5 | $test(atX, \ldots)$, $atX$ true | $(\ell_0^1, (2,0), 2,0)$ |
| 6 | $shift(E, \ell_0^3)$ | $(\ell_0^3, (2,0), 3,0)$ |
| 7 | $test(atO, \ldots)$, $atO$ false | $(\ell_0^4, (2,0), 3,0)$ |
| 8 | $move(W, \ell_0^3)$ | $(\ell_0^3, (1,0), 3,0)$ |
| 9 | $test(atO, \ldots)$, $atO$ false | $(\ell_0^4, (1,0), 3,0)$ |
| 10 | $move(W, \ell_0^3)$ | $(\ell_0^3, (0,0), 3,0)$ |
| 11 | $test(atO, \ldots)$, $atO$ true | $(q_{\ell_h}, (0,0), 3,0)$ |

The final configuration is $E_P(\ell_h, 3,0)$: the machine halts with marker $X$ at $(3,0)$ and the cursor back at the origin, so the output is 3. The single CM2-instruction is realized by 11 primitive navigation steps, in accordance with the count $4m + 3$ established in the proof of Theorem 4.5.

Branching is exercised by $P_{tr} = (\{\ell_0, \ell_1, \ell_h\}, \iota)$ with $\iota(\ell_0) = jzdec(1, \ell_h, \ell_1)$ and $\iota(\ell_1) = inc(2, \ell_0)$, which transfers $c_1$ into $c_2$. On input $(m, n) = (2,0)$ the CM2-computation is

$$(\ell_0, 2,0) \vdash (\ell_1, 1,0) \vdash (\ell_0, 1,1) \vdash (\ell_1, 0,1) \vdash (\ell_0, 0,2) \vdash (\ell_h, 0,2),$$

and by Theorem 4.5 the corresponding navigation computation consists of five macro-steps, each beginning and ending with the cursor at the origin: two east-walk-and-decrement macros on the horizontal axis (taking the false branch of the zero test), two north-walk-and-increment macros on the vertical axis, and a final zero test that succeeds without moving anything and passes control to $q_{\ell_h}$. Both branches of the conditional and both axes are thus used.

This first proof is mathematically straightforward and historically unsurprising, but it makes vivid the idea that two allocentrically distinguished axes can function as two unbounded spatial registers.

## 5. Direct universality via a writable tape-path in the map

The second navigation model makes the simulation more direct. Instead of encoding memory arithmetically as distances of markers from the origin, it places writable labels on landmarks along a distinguished bi-infinite path through the allocentric map. The current simulated location plays the role of the Turing head, and movement along the path plays the role of left and right head motion. In effect, this model is an isomorphic recoding of an ordinary one-tape Turing machine into allocentric-map terms: the path realizes the tape, the current simulated location realizes the head position, and the stored labels on path landmarks realize the tape contents.

**Definition 5.1 (Tape-path allocentric navigation machine).** Fix a finite alphabet $\Gamma$ with blank $\square \in \Gamma$. Let $\Lambda_\Gamma = \{\lambda: \mathbb{Z}^2 \to \Gamma \mid \lambda(z) = \square$ for all but finitely many $z \in \mathbb{Z}^2\}$ be the set of landmark labelings: like $T_\Gamma$, it is a set of finite-support functions, from which each configuration draws one member. Let

$$\pi: \mathbb{Z} \to \mathbb{Z}^2, \pi(i) = (i, 0),$$

and write $L_\pi = \pi(\mathbb{Z})$ for the set of path landmarks. Note that $\pi$ is injective and that $\pi(i \pm 1)$ is a four-neighbor of $\pi(i)$ for every $i \in \mathbb{Z}$.

A tape-path allocentric navigation machine is a tuple

$$A = (\mathbb{Z}^2, \pi, Q, \Gamma, \square, q_0, q_h, \delta_A), \quad \delta_A: (Q \backslash \{q_h\}) \times \Gamma \to Q \times \Gamma \times \{pred, succ\}.$$

A configuration is a triple $(q, c, \lambda) \in Q \times L_\pi \times \Lambda_\Gamma$, where $c$ is the current simulated location, constrained to lie on the path. The one-step relation is: with $a = \lambda(c)$ and $\delta_A(q, a) = (q', b, \mu)$,

$$(q, c, \lambda) \Rightarrow_A (q', c', \lambda[c \mapsto b]),$$

where $c' = \pi(i-1)$ if $\mu = pred$ and $c' = \pi(i+1)$ if $\mu = succ$, $i$ being the unique index with $c = \pi(i)$.

Configurations with control state $q_h$ are terminal; all others have exactly one successor.

In one computational step the machine therefore checks the stored label at the current path location, replaces that label, changes control state, and moves to the predecessor or successor location on the path. The path is unbounded in the same sense as a Turing tape, and by Convention 3.8 a concrete configuration represents only finitely many nonblank labels.

**Remark 5.2 (On the choice of embedding).** Any injection $\pi': \mathbb{Z} \to \mathbb{Z}^2$ such that $\pi'(i \pm 1)$ is a four-neighbor of $\pi'(i)$ for every i—for example, the vertical embedding $\pi'(i) = (0, i)$—yields a computationally equivalent tape-path model under the relabeling $\pi' \circ \pi^{-1}$ of path locations. The geometric locus of the simulated tape therefore carries no mathematical weight. The substantive assumption is that the path's adjacency structure is handed to the machine as a primitive: $pred$ and $succ$ are defined by reference to $\pi$. Section 6 removes that primitive.

**Definition 5.3 (Translation and encoding).** Let $tr(L) = pred$ and $tr(R) = succ$. Given a one-tape Turing machine $M = (Q, \Gamma, \square, q_0, q_h, \delta)$, define

$$A_M = (\mathbb{Z}^2, \pi, Q, \Gamma, \square, q_0, q_h, \delta^{tr}),$$

where $\delta^{tr}(q, a) = \big(q', b, tr(D)\big)$ whenever $\delta(q, a) = (q', b, D)$.

The map $M \mapsto A_M$ is computable: it is the identity on states, alphabet, blank, start state, and halt state, and it relabels the third component of each entry of the transition table. Define the encoding

$$E_M: Conf(M) \to Conf(A_M), \quad E_M(q,h,t) = (q, \pi(h), \lambda_t),$$

where $\lambda_t(\pi(i)) = t(i)$ for every $i \in \mathbb{Z}$ and $\lambda_t(z) = ▫$ for every $z \notin L_\pi$. Since $t \in T_\Gamma$ has finite support, $\lambda_t \in \Lambda_\Gamma$; and $t \mapsto \lambda_t$ is injective, so $E_M$ is injective. The notation $\lambda_t$ is a compact metatheoretic name for a finite-support labeling; it is not an infinite object supplied to the machine.

**Theorem 5.4.** For every deterministic one-tape Turing machine $M$, the encoding $E_M$ of Definition 5.3 is a lock-step simulation of $M$ in $A_M$. Consequently, the class of tape-path allocentric navigation machines is Turing-complete.

*Proof.* Terminality: $\mathbf{A_M}$-configurations with state $q_h$ are $\Rightarrow$-terminal and $M$-configurations with state $q_h$ are $\vdash$-terminal, and $\delta^{tr}$ is defined exactly where $\delta$ is; this gives condition (b) of Definition 3.5. For condition (a), take any non-terminal configuration $\gamma = (q,h,t)$ of $M$, let $a = t(h)$ and $\delta(q,a) = (q', b, D)$, so that $\gamma \vdash_M \gamma' = (q', h', t[h \mapsto b])$ with $h' = h - 1$ if $D = L$ and $h' = h + 1$ if $D = R$. Now $E_M(\gamma) = (q, \pi(h), \lambda_t)$ and $\lambda_t(\pi(h)) = t(h) = a$, so the machine $A_M$ applies $\delta^{tr}(q,a) = (q', b, tr(D))$. It writes $b$ at $\pi(h)$, yielding the labeling $\lambda_{t[\pi(h)\mapsto b]} = \lambda_{t[h\mapsto b]}$ — the two labelings agree at every path landmark by construction and are both blank off the path — changes state to $q'$, and moves to $\pi(h-1)$ if $tr(D) = pred$ and to $\pi(h+1)$ if $tr(D) = succ$. The result is $(q', \pi(h'), \lambda_{t[h\mapsto b]}) = E_M(\gamma')$. Since $k = 1$, no intermediate configuration arises and the simulation is lock-step. Because ordinary one-tape Turing machines are universal, the class is Turing-complete. ■

Relative to the first proof, this model is more direct and conceptually closer to standard Turing-machine organization. Relative to older graph-based formalisms, however, it remains a familiar kind of tape-on-graph encoding.

**Corollary 5.5.** Let $U$ be a universal one-tape Turing machine, of which many small examples are known [27, 49]. Then $A_U$ is a single tape-path allocentric navigation machine that simulates every one-tape Turing machine, given an encoding of that machine and its input on the path.

*Proof.* Immediate from Theorem 5.4 applied to $U$, together with the universality of $U$. ■

Corollary 6.8 below strengthens this to universality for the class itself.

## 6. Direct universality via a fully local two-dimensional landmark layout

The third navigation model removes the globally designated path from the machine's resources. Memory is realized as a two-dimensional field of landmarks, and the machine moves only by following local predecessor/successor information available at the landmark it currently occupies. Equivalently, the model gives a local two-dimensional embedding of a linear Turing tape into a zigzag track in the allocentric map: the machine itself accesses only local direction data, while the linear order appears only in the metatheory.

**Definition 6.1 (Local-layout allocentric navigation machine).** Let the internal map be $\mathbb{Z}^2$ and let the memory landmarks occupy the two-row track

$$T = \{(x, y) \in \mathbb{Z}^2 \mid y \in \{0,1\}\}.$$

Define the landmark type function $\tau: T \to \{A, B, C, D\}$ by: $\tau(x, y) = A$ if $y = 0$ and $x$ is even; $B$ if $y = 1$ and $x$ is even; $C$ if $y = 1$ and $x$ is odd; $D$ if $y = 0$ and $x$ is odd. Identify each direction with a unit vector: $E = (1,0), W = (-1,0), N = (0,1), S = (0, -1)$. Define the two local direction dictionaries $pred_dir, succ_dir: \{A, B, C, D\} \to \{E, W, N, S\}$ as listed in Table 2.

*Table 2 Local predecessor and successor directions by landmark type*

| type | occupies | pred_dir | succ_dir |
|---|---|---|---|
| A | (2k, 0) | W | N |
| B | (2k, 1) | S | E |
| C | (2k+1, 1) | W | S |
| D | (2k+1, 0) | N | E |

A local-layout allocentric navigation machine is a tuple

$$A = (\mathbb{Z}^2, T, \tau, Q, \Gamma, \,\square\, , q_0, q_h, \delta_A), \quad \delta_A: (Q \backslash \{q_h\}) \times \Gamma \to Q \times \Gamma \times \{pred, succ\}.$$

A configuration is a triple $(q, c, \lambda) \in Q \times T \times \Lambda_\Gamma$, where $c$ is the landmark bearing the head marker and $\lambda$, as in Definition 5.1, is a finite-support labeling, here supported in $T$. The one-step relation is: with $a = \lambda(c)$ and $\delta_A(q, a) = (q', b, \mu)$,

$$(q, c, \lambda) \Rightarrow_A (q', c + v, \lambda[c \mapsto b]),$$

where $v = pred_dir\big(\tau(c)\big)$ if $\mu = pred$ and $v = succ_dir\big(\tau(c)\big)$ if $\mu = succ$.

Configurations with control state $q_h$ are terminal; all others have exactly one successor.

The landmarks are thereby arranged in a zigzag chain through the plane:

$$\cdots \to (-2,0) \to (-2,1) \to (-1,1) \to (-1,0) \to (0,0) \to (0,1) \to (1,1) \to (1,0) \to (2,0) \to \cdots$$

**Remark 6.2 (What is and is not available to the machine).** The type function $\tau$ is part of the machine's readable environment: at each step, the next location is determined by the type of the landmark currently occupied, together with the fixed dictionaries $pred_dir$ and $succ_dir$ in finite control. The machine has no access to any other global description of the layout. In particular, it has no path object or landmark index; the enumeration $\eta$ in Lemma 6.3 is used only in the correctness proof. Removing the landmark types would leave the machine with no move relation, because the geometric direction of $pred$ and $succ$ is defined nowhere else. Relative to Definition 5.1, what this model eliminates is the globally indexed path $\pi$ whose adjacency relation is primitive for the tape-path machine. This is why the local-layout model is not merely the tape-path model re-embedded in the plane, as the variant described in Remark 5.2 is.

**Lemma 6.3.** Define $\eta: \mathbb{Z} \to T$ by

$$\eta(4k) = (2k, 0), \quad \eta(4k+1) = (2k, 1), \quad \eta(4k+2) = (2k+1, 1), \quad \eta(4k+3) = (2k+1, 0), \quad k \in \mathbb{Z}.$$

Then $\eta$ is a bijection and, for every $i \in \mathbb{Z}, \eta(i) + succ_dir\left(\tau(\eta(i))\right) = \eta(i+1)$ and $\eta(i) + pred_dir\left(\tau(\eta(i))\right) = \eta(i-1)$.

*Proof.* For each $k$, the four values $\eta(4k), \dots, \eta(4k+3)$ are $(2k, 0), (2k, 1), (2k+1, 1), (2k+1, 0)$, so as $k$ ranges over $\mathbb{Z}$ every element of $T$ is hit exactly once; $\eta$ is a bijection. For the successor claim, take the four residues in turn. If $i = 4k$, then $\eta(i) = (2k, 0)$ has type A, so $succ_dir = N$ and $\eta(i) + N = (2k, 1) = \eta(4k+1)$. If $i = 4k+1$, then $\eta(i) = (2k, 1)$ has type B, so $succ_dir = E$ and $\eta(i) + E = (2k+1, 1) = \eta(4k+2)$. If $i = 4k+2$, then $\eta(i) = (2k+1, 1)$ has type C, so $succ_dir = S$ and $\eta(i) + S = (2k+1, 0) = \eta(4k+3)$. If $i = 4k+3$, then $\eta(i) = (2k+1, 0)$ has type D, so $succ_dir = E$ and $\eta(i) + E = (2k+2, 0) = \eta\big(4(k+4)\big) = \eta(i+1)$. The predecessor claim is verified by the same four cases: from type A, $pred_dir = W$ and $(2k, 0) + W = (2k-1, 0) = \eta(4k-1)$; from type B, $(2k, 1) + S = (2k, 0) = \eta(4k)$; from type C, $(2k+1, 1) + W = (2k, 1) = \eta(4k+1)$; from type D, $(2k+1, 0) + N = (2k+1, 1) = \eta(4k+2)$. Hence following the local pointers coincides exactly with increment and decrement of the index $i$. ■

**Definition 6.4 (Translation and encoding).** With $tr$ as in Definition 5.3, define for each one-tape Turing machine $M$ the local-layout machine

$$A'_M = (\mathbb{Z}^2, T, \tau, Q, \Gamma, ▫, q_0, q_h, \delta^{tr}),$$

and the encoding

$$E'_M: Conf(M) \to Conf(A'_M), \quad E'_M(q, h, t) = (q, \eta(h), \lambda_t),$$

where $\lambda_t\big(\eta(i)\big) = t(i)$ for every $i \in \mathbb{Z}$ and $\lambda_t(z) = ▫$ for $z \notin T$. As in Definition 5.3, $\lambda_t \in \Lambda_\Gamma$ and $E'_M$ is an injection. The order of the components matches Definition 5.3: control state, current location, labeling.

**Theorem 6.5.** For every deterministic one-tape Turing machine $M$, the encoding $E'_M$ of Definition 6.4 is a lock-step simulation of $M$ in $A'_M$. Consequently, the class of local-layout allocentric navigation machines is Turing-complete.

*Proof.* Terminality is as in Theorem 5.4. For the step condition, let $\gamma = (q, h, t)$ be non-terminal, $a = t(h)$, $\delta(q, a) = (q', b, D)$, and $\gamma \vdash_M \gamma' = (q', h', t[h \mapsto b])$ with $h' = h - 1$ if $D = L$ and $h' = h + 1$ if $D = R$. Then $E'_M(\gamma) = (q, \eta(h), \lambda_t)$ and $\lambda_t\big(\eta(h)\big) = t(h) = a$, so $A'_M$ applies $\delta^{tr}(q, a) = \big(q', b, tr(D)\big)$. It writes $b$ at $\eta(h)$, giving $\lambda_{t[\eta(h) \mapsto b]} = \lambda_{t[h \mapsto b]}$, changes state to $q'$, and moves by the vector $pred_dir\left(\tau(\eta(h))\right)$ or $succ_dir\left(\tau(\eta(h))\right)$ according to $tr(D)$. By Lemma 6.3 that move reaches $\eta(h-1)$ in the $pred$ case and $\eta(h+1)$ in the $succ$ case, that is, $\eta(h')$. The result is $E'_M(\gamma')$, and the simulation is lock-step. Because ordinary one-tape Turing machines are universal, the class is Turing-complete. ■

This third proof is the most faithful to the idea of computation by navigation among landmarks, because the resources the machine operates with are entirely local. In that respect it lies especially close, in mathematical spirit, to local graph-based machine models studied in computability theory, even though it is recast here in allocentric cognitive-map terms.

**Remark 6.6 (On the word "symbol").** In philosophy of mind and the mind sciences, "symbol" often carries commitments absent in computability theory. The proofs of Theorems 5.4 and 6.5 assign locations within a cognitive map formal roles corresponding to entries in the tape alphabet $\Gamma$, but this does not imply that allocentric navigation requires symbols in the sense of having language-like structure, or even in the weaker sense of having arbitrary semantic content. Letters of a finite alphabet are just reliably distinguishable labels; nothing in the proofs depends on stored map items being symbolic in a richer sense. To make this explicit, let $\Omega$ be a finite set of reliably distinguishable stored item-types and let $\rho: \Omega \to \Gamma$ be a fixed bijection. Each relevant map location stores an item of type $\omega \in \Omega$, realizing the tape letter $\rho(\omega)$. Thus, the map need only store one of finitely many reliably distinguishable item-types, not literal elements of $\Gamma$. This relabeling changes no mathematics; it blocks only a misleading identification of finitely many labels with symbols in stronger senses. Defenders of the classical Language of Thought hypothesis have argued that architectures without language-like symbols lack the power to explain cognition. The proofs above assume no such symbols. This is one respect in which the navigation-based neurocomputational architectures explored here are nonclassical.

**Corollary 6.7 (Turing equivalence).** Under Convention 3.8, every machine in each of the three navigation-model classes is simulable by an ordinary Turing machine. Therefore, the three classes are Turing-equivalent to standard universal models.

*Proof.* In the counter-based model, a configuration is the finite tuple $(q, p, x, y)$, and each primitive test and update of Definition 4.2 is a computable operation on natural numbers, monus included. In the tape-path model, a configuration is represented by the finite control state, the current path index or coordinate, and the finite list of nonblank path labels; $\pi$ and its inverse on $L_\pi$ are computable. In the local-layout model, a configuration is represented by the finite control state, the current landmark coordinate, and the finite list of nonblank landmark labels; $\tau$ is computable from the parity of the first coordinate and the value of the second, and $pred_dir$ and $succ_dir$ are finite tables. In each case a standard Turing machine can store the finite description of the machine and the finite representation of the current configuration, compute the next primitive step, and iterate. No infinite completed function, oracle, or non-effective operation is added. Combined with Theorems 4.5, 5.4, and 6.5, this yields Turing equivalence. ■

**Corollary 6.8 (Universal navigation machines).** Each of the three classes of navigation machines contains a machine that is universal for that class in the sense of Definition 3.7.

*Proof.* First let $K$ be the tape-path class or the local-layout class, and let $A \in K$ be arbitrary. By Corollary 6.7 there exist a Turing machine $M_A$ and computable codings such that $M_A$ simulates $A$; the construction there is uniform in $A$, so the map $\langle A \rangle \mapsto \langle M_A \rangle$ is computable. Let $U$ be a universal one-tape Turing machine. Then $U$ simulates $M_A$ given $\langle M_A \rangle$ and an encoded input, and by Theorem 5.4 (respectively Theorem 6.5) the machine $A_U \in K$ simulates $U$ lock-step. Composing the three simulations and the corresponding computable codings, $A_U$ on input $\langle A \rangle$ followed by any input $w$ for $A$ computes what $A$ computes on $w$. Since $A \in K$ was arbitrary, $A_U$ is universal for $K$.

For the counter-based class the composition differs at its middle step. Let $A$ be an arbitrary counter-based allocentric navigation machine and let $M_A$ be as above. By Remark 3.3 there is a CM2-program $P_U$ that, given a Gödel-coded description of an arbitrary one-tape Turing machine together with a coded input, computes what that machine computes; and by Theorem 4.5 the machine $A_{P_U}$ simulates $P_U$ step for step. Composing, $A_{P_U}$ computes what $A$ computes, given $\langle M_A \rangle$ together with a coded input. Since $A$ was arbitrary, $A_{P_U}$ is universal for the counter-based class, under the input convention noted in Remark 3.3. ∎

Corollary 6.8 matters for the interpretation of the results. It is not merely that for each Turing machine there exists some navigation machine that does what it does. There is a single, fixed navigation machine that, supplied with a description of an arbitrary machine of its class and an input, computes what that machine computes. The word "universal" in the title of this article is therefore warranted in both of the senses distinguished in Definitions 3.6 and 3.7.

## 7. Comparative significance

The three theorems establish the same computability-theoretic conclusion while differing in how they apply navigation-based processing. Together with Corollaries 6.7 and 6.8, they show Turing equivalence under ordinary finitary assumptions and the existence of universal machines within each class. The result is that the representational and operational primitives of allocentric navigation are sufficient, once idealized, to realize the same class of computable transformations as standard symbolic models. This should not be mistaken for a novel or more powerful kind of Turing-style machine; as with familiar Turing-machine variants, changes in storage format can preserve ordinary computable power [50].

The counter proof is the most austere. It uses two one-dimensional axes embedded in a common allocentric frame rather than a fully exploited two-dimensional map. The orthogonality of the axes separates the two unbounded stores, while marker position encodes counter value. Its distinctive resource is the locality constraint of Definition 4.2(iii): the machine cannot touch a marker without first traveling to it, which is why each counter instruction compiles into a macro whose length grows with the counter value. In this respect the proof is closer to register-machine computation than to ordinary wayfinding in a two-dimensional environment.

The tape-path proof is more direct. In fact, it is essentially an isomorphic recoding of an ordinary one-tape Turing machine into an allocentric cognitive map. In the broad sense of “map” used in this article, a Turing-machine tape is already a one-dimensional map: it is an ordered array of locations, each of which can store one of finitely many item-types. If a navigation architecture can maintain a track of landmarks with replaceable stored items and move a simulated location along that track, then the standard Turing model is already present almost verbatim. The advantage we gain here is interpretive clarity: landmarks are memory cells, their stored items are tape entries, and navigation is head movement. The disadvantage is that the proof still presupposes a globally distinguished path whose adjacency relations are simply handed to the machine.

The local-layout proof is more local and spatially natural. It may be viewed as a local two-dimensional embedding of a linear Turing tape into a field of landmarks. The memory is realized as a two-dimensional field of landmarks, and the machine moves only by inspecting local direction data at its present position. The hidden indexing used in the proof exists only at the meta-level to establish correctness. Thus, the resources the machine operates with are fully local even though the proof uses a global enumeration for convenience.

It is worth being explicit about what does and does not distinguish the three models. They are not distinguished by where in the plane the simulated memory is placed. A tape-path model built from any other four-neighbor embedding of the same bi-infinite path is computationally equivalent to the model of Section 5, as Remark 5.2 notes. What distinguishes the models is which resources the machine is granted: in Section 4, two axis-bound markers, three tests, and no writable labels; in Section 5, writable labels plus a path whose adjacency relation is a primitive of the model; in Section 6, writable labels plus nothing but locally readable direction data, with no path object in the model at all. That third difference is not a difference of presentation. The move relation of Definition 6.1 is definable from the machine's local resources alone, whereas the move relation of Definition 5.1 is not.

The two-dimensional grid in the third model is not essential from the standpoint of computability theory. Any graph containing a bi-infinite simple path—or, for a particular finite computation, a simple path long enough to contain the portion of tape it visits—together with the appropriate writable local memory and local predecessor/successor cues can support the same type of simulation. The reason for using a two-dimensional allocentric layout is not mathematical necessity but architectural interpretation: it keeps the formal model close to the language of landmarks, directions, and local navigation within a map.

A separate issue concerns finite control. The models do require a finite internal state and a finite transition table. That component is not itself a map-like memory store; it is the finite control component familiar from ordinary Turing machines, counter machines, graph Turing machines, and storage-modification machines. The claim that the models are map-based is therefore not the claim that everything in the computation is literally stored in the map. It is the claim that the unbounded memory, the accessed items, and the head or cursor movements are realized in map-like terms. In that respect, the models resemble ordinary graph-search algorithms such as Dijkstra's algorithm and A*, which also combine a graph or map with finite control and update rules [51, 52]. The difference is that the present models use writable landmark memory to perform arbitrary effective computation, not merely shortest-path search.

At the neurocognitive level, this finite-control component is where prefrontal control systems are most relevant. Prefrontal cortex (PFC) and related orbitofrontal (OFC) and cingulate systems provide goal, rule, and policy signals that regulate which map states are accessed, which transitions are selected, and whether navigation is continued, inhibited, or replanned. This is the kind of architectural role played by finite control acting through the head in classical machines: a limited control resource coordinates operations over a memory substrate. This analogy is consistent with general accounts of PFC as cognitive control [53] and with evidence that prefrontal systems contribute to both online navigation and offline navigational simulation: prefrontal-thalamic circuits are required for hippocampal route representations in goal-directed spatial navigation [18], OFC represents future navigational goals [16], hippocampal goal representations in human route planning covary with goal-related coding in PFC [17], medial PFC perturbation alters hippocampal sequences during deliberative decision making [20], and medial PFC replay can represent generalized trajectories during rule switching [19, 54].

This role for transitions among states or locations is not foreign to cognitive-map theorizing. Contemporary work often treats cognitive maps as structures that support transitions, planning, and prediction over states or locations, including successor or predictive-map models and more general accounts of cognitive maps as structured representations for flexible behavior [5, 6, 13, 55]. The present models extend, generalize, and idealize those ideas: their transitions are exact, their labels are perfectly writable, and their maps are unbounded. But the abstract ingredients—locations, local transitions, stored information, and finite control—are recognizable extensions of resources already invoked in the literature on cognitive maps.

It might be objected that questions of this kind are already settled by the observation that neural networks are computationally universal, so that the architecture at issue must be a network architecture. This objection fails to distinguish three issues. First, neural-network models are computationally universal only when supplied with some unbounded computational resource, such as indefinitely extensible memory or, in some finite recurrent models, arbitrarily precise internal states [e.g., 56]. A fixed finite network with only finitely many distinguishable internal states cannot realize unbounded computation and therefore is not computationally universal. Second, even if some neural-network model is universal, that fact by itself says nothing about which architecture the brain realizes. Third, the foundational disputes mentioned in Section 1 concern architecture at the level of maps, landmarks, and navigational operations, not network or circuit organization [28, 29]. A general-purpose network is compatible with many neurocomputational architectures, and pointing to a network formalism does not identify which architecture biological cognition uses. The present results concern the power of one empirically motivated architecture, described in the vocabulary used by the relevant neuroscience.

The results of this article have implications for the claim that navigation-based or otherwise nonclassical architectures lack sufficient representational or computational power to support cognition. This sort of objection depends on an additional assumption, namely that adequate representation must take a symbolic, language-like form. The results developed here undermine that assumption. They show that systems grounded in structured spatial relations and local transition rules can realize the same class of computable transformations typically associated with symbolic architectures — and they do so from resources that such critics have tended to treat as noncomputational, as merely implementational, or as barely cognitive: what they have taken to be the neural machinery of getting from one place to another.

The point extends to stored-program organization, which might seem the strongest case for the other side (the models of Sections 4 through 6 do not require it; see Definition 4.1). Even if a navigation architecture supported it, classical symbolic architecture would not thereby be vindicated; what would follow is that the map-based resources are sufficient to realize it. The more such resources turn out to support, the less need there is to posit a separate symbolic architecture alongside them. Accordingly, symbolic processing itself could be understood as a special case of spatial navigation, where symbols are items represented at locations within a cognitive map and symbolic operations are actions taken on such items. Developing this suggestion in detail is a research program I pursue in separate work.

## 8. Limitations and scope

These universality theorems require standard idealizations. They assume an unbounded allocentric map, perfectly reliable landmark identity, exact storage, exact local movement, and arbitrarily long navigation. None of those assumptions should be attributed to biological systems. Therefore, these results do not establish that animals, humans, or hippocampal-entorhinal systems ordinarily implement Turing-universal navigation machines. They establish only that allocentric navigation-based computational models, if idealized in the usual way familiar from computability theory, are computationally universal.

The unbounded maps used here should be understood as potential infinities, not as completed infinite resources. This is the role played by the tape in the ordinary Turing-machine model. At any finite time in any finite computation, only finitely many locations have been visited and only finitely many nonblank labels or marker positions must be represented. The constructions therefore remain within the finitary conception of effective computation.

The evidence from neuroscience supports allocentric spatial representation, place and grid coding, and map-like navigation capacities, including offline navigation. It does not by itself support the stronger assumptions required for universal computation, such as exact unbounded writable memory at landmarks. The contribution is thus a formal reconstruction that begins from an empirically motivated representational vocabulary and then idealizes it into a universal machine model.

The reconstruction assumes that a navigational map is a structure on which operations are performed and through which action is organized, rather than only a device for finding or choosing paths. The evidence supports that reading. Hippocampal-entorhinal map-like codes are recruited for memory, inference, and abstract relational structure as well as for wayfinding [11-14], and hippocampal population activity during behavior tracks the animal's evolving goal and action plan rather than only sensory or spatial variables [15].

These maps are also used offline: replay may support consolidation and retrieval, and preplay can express sequences related to future navigation [9, 10]. The primitives idealized in Sections 4 through 6—locations, items at locations, movement among them, and operations performed at them—are drawn from that broader, action-oriented picture.

The theorems also do not show that every navigation architecture is universal. Universality depends on specific structural enrichments: unbounded storage, reliable update operations, and sufficiently precise control. A purely finite navigation system, or a system lacking sufficiently stable memory, could not be strictly universal. However, the navigation and action processes discussed here may also occur partly within the external environment, which under suitable idealization may be treated as an external memory store. External media including paper, diagrams, or other items can in principle supply some of the structure used by a navigation system. In that case, the same universal computational models can be realized by allocentric navigation architectures that distribute storage across both internal and environmental resources. This observation is relevant to the architecture of human mathematical calculation: the same kinds of mapping and control resources involved in allocentric navigation may be involved in mathematical calculation and may help organize the spatially structured external memory used in such tasks.

Explaining human calculational competence requires additional neurocognitive assumptions. Miguel-Tomé's 2024 preprint, *The Numeric Map Principle: A Navigational-Based Hypothesis to Explain the Human Skills to Perform and Understand Basic Arithmetic* (https://doi.org/10.5281/zenodo.13921658), addresses that more restricted question directly. It hypothesizes that the human navigational system can construct a one-dimensional numeric map isomorphic to a fragment of arithmetic and thereby help ground basic whole-number arithmetic in navigational operations. That hypothesis is complementary to the present results and independent of them. My target here is more abstract and general: I use standard computability-theoretic simulations to ask whether idealized allocentric-map computational models can realize arbitrary Turing-computable transformations. I do not claim that allocentric navigation by itself explains human mathematical understanding, and nothing in the argument presupposes the Numeric Map Principle. The claim is that map-based architectures have enough computational power in principle to support arbitrarily complex computable cognition.

The present results concern the computational power of idealized allocentric navigation architectures once their operative map structure, landmark relations, and transition resources are fixed. Under such conditions, allocentric navigation is Turing-computable in the ordinary sense. This does not imply that all biological processes involved in map formation, map revision, exploratory behavior, learning, or environmental interaction are themselves Turing-computable or fully capturable as computational processes. Real biological navigation unfolds in open-ended interaction with changing environments and dynamically evolving neural and bodily systems. I do not attempt to characterize the full dynamics of such processes here. My more limited claim is that allocentric navigation architectures possess sufficient representational and operational resources to realize universal computation under standard computability-theoretic idealizations.

## 9. Conclusion

I have presented three formal routes from allocentric cognitive-map architectures to computational universality. The first uses two allocentrically distinguished axes to encode counter values. The second turns a distinguished path of writable landmarks into a direct analogue of a Turing tape. The third replaces that global path with a fully local two-dimensional layout whose predecessor and successor relations are available only through locally stored landmark structure. Each model has been specified as a tuple with an explicit instruction syntax and one-step relation, each simulation proceeds by a translation function together with an explicit encoding of configurations, and universality has been established both for the classes and, within each class, for a machine that simulates every machine in it.

These results are mathematically continuous with earlier work on graph-based computation, but they make explicit how universal computational capacity can be reconstructed within the framework of navigation within allocentric cognitive maps. With the finitary restrictions made explicit above, the three model classes are Turing-equivalent to standard universal models. In that way, they clarify the representational and computational significance of a class of architectures that are independently motivated by empirical work in neuroscience.

None of this is to say that biological navigation literally implements classical Turing-style computation. On the contrary, if allocentric navigation—which is nonclassical in many respects—is expressive enough to implement classical Turing computation, it may be expressive enough to explain cognition nonclassically (as I argue more generally in *Neural Hardware for the Language of Thought: New Rules for an Old Game*; https://arxiv.org/abs/2510.10251). That is enough to motivate the research program mentioned in Section 7, in which higher cognition, including symbolic processing, is reconstructed in navigation-based terms.

## Statements and Declarations

**Funding.** No funding was received to assist with the preparation of this manuscript.

**Competing interests.** The author has no relevant financial or non-financial interests to disclose.

**Data availability.** No datasets were generated or analyzed during the current study.

**Use of generative AI.** I used ChatGPT, Google Gemini, and Claude for research assistance, drafting and checking proofs, feedback on drafts, and editing. I reviewed and revised all output. I am solely responsible for all substantive ideas, arguments, interpretations, final source selection, and final wording.